# Microscopic dynamics of charge separation at the aqueous electrochemical interface

John A. Kattirtzi
*Department of Chemistry, Massachusetts Institute of Technology, Cambridge, MA 02138, USA and*
*College of Chemistry and Chemical Engineering,*
*Xiamen University, Xiamen 361005, People's Republic of China*

David T. Limmer
*Department of Chemistry, University of California, Berkeley, CA 94609, USA*
*Kavli Energy NanoScience Institute, University of California, Berkeley, CA 94609, USA and*
*Material Science Division, Lawrence Berkeley National Laboratory, Berkeley, CA 94609, USA*

Adam P. Willard
*Department of Chemistry, Massachusetts Institute of Technology, Cambridge, MA 02138, USA*
(Dated: March 6, 2017)

We have used molecular simulation and methods of importance sampling to study the thermodynamics and kinetics of ionic charge separation at a liquid water-metal interface. We have considered this process using canonical examples of two different classes of ions: a simple alkali-halide pair, $Na^+I^-$, or classical ions, and the products of water autoionization, $H_3O^+OH^-$, or water ions. We find that for both ion classes, the microscopic mechanism of charge separation, including water's collective role in the process, is conserved between the bulk liquid and the electrode interface. Despite this, the thermodynamic and kinetic details of the process differ between these two environments in a way that depends on ion type. In the case of the classical ion pairs, a higher free energy barrier to charge separation and a smaller flux over that barrier at the interface, results in a rate of dissociation that is 40x slower relative to the bulk. For water ions, a slightly higher free energy barrier is offset by a higher flux over the barrier from longer lived hydrogen bonding patters at the interface, resulting in a rate of association that is similar both at and away from the interface. We find that these differences in rates and stabilities of charge separation are due to the altered ability of water to solvate and reorganize in the vicinity of the metal interface.

In aqueous solution, the association or dissociation of oppositely charged ions requires the collective rearrangement of surrounding water molecules [1, 2], which solvate bound ion pairs differently than individual ions [3–6]. The solvent fluctuations that enable these collective rearrangements drive the dynamics of ion pairing and unpairing and therefore play a fundamental role in many chemical reactions. Near the surface of an electrode these collective solvent fluctuations can differ significantly from that of the bulk liquid [7, 8] and these differences can affect the rates and mechanisms of aqueous electrochemical reactions. In this manuscript, we use molecular simulation to investigate the microscopic processes of aqueous ion pairing when it takes place near but not in direct contact with an extended metallic electrode. We identify the specific effects of the electrode interface by comparing our results to those generated in the environment of the bulk liquid. We find that the presence of an electrode has little effect on the mechanistic details of ionic charge separation, but can significantly influence the thermodynamics and kinetics of the process. We highlight that this influence is different for simple monovalent salts like $Na^+$ and $Cl^-$, whose transport is limited by the mobility of an aqueous solvation shell [9], than it is for water ions, like $H_3O^+$ and $OH^-$, whose transport is limited by the concerted hopping of protons along hydrogen bonding chains [10]. This fundamental difference is controlled by the microscopic details of electrode-water interactions, and thus has implications for the nanoscale design of aqueous electrochemical systems.

When described in terms of a chemical reaction,

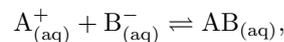
$$A^+_{(aq)} + B^-_{(aq)} \rightleftharpoons AB_{(aq)},$$

the process of aqueous ion association (forward reaction above) or dissociation (reverse reaction above) is deceptively simple. The expression omits the cooperative role of solvent, which must restructure to accommodate transitions between the associated and dissociated states [2, 11]. This restructuring, which is driven by thermal fluctuations, is both collective (extending beyond the first solvation shell) and fleeting [12–14]. These microscopic processes are difficult to probe experimentally[15] so atomistic simulation has played an important role in revealing their molecular-level details. Simulations based on state of the art quantum chemical methods reveal important information about the structure and energetics of these systems, however, without additional importance sampling or embedding, they are too computationally demanding to characterize collective room temperature dynamics[16–19]. Classical simulations overcome this limitation by treating some interactions empirically, thereby enabling the characterization of the equilibrium dynamics of extended molecular systems. The tools of modern statistical mechanics, combined with computational techniques for efficiently sampling reactive trajectories[20], enabled a detailed characterization of the reaction mechanism, thermodynamics, and kinetics of these processes. These tools have been previously applied to investigate ion pair dissociation in the bulk liquid, which has revealed that the dissociation of classical ions and water ions are both crucially driven by electrostatic fluctuations of the aqueous environment [2, 6]. The characteristics of these electrostatic fluctuations are similar between the two types of ion pairs despite the fact that their solvation structures differs significantly.

When simple classical ions are solvated in aqueous solution, they are dressed by a solvation shell of water molecules whose orientations are polarized in response to the ionic charge. Similarly, bound pairs of ions are dressed by water molecules whose orientations are polarized in response to the electric dipole of the bound ion pair. In order for a bound pair of ions to separate, this dipolar solvation shell must be deconstructed and transformed into two separate and oppositely polarized ionic solvation shells. This solvent reorganization has been identified as the rate-limiting step for aqueous ion dissociation, and numerous efforts have been aimed at quantifying water's role in this process [1–4, 14]. These efforts have revealed that although the overall process of aqueous ion dissociation is usually thermodynamically favorable, solvent reorganization leads to the emergence of a free energy barrier that is on the order of typical thermal energies, $k_B T$, where $k_B$ is Boltzmann's constant and $T$ is the temperature.

Water ions, specifically hydroxide and hydronium, are the products of proton transfer to and from an individual water molecule. These ions can easily integrate into the aqueous hydrogen bonding network and leverage the Grotthuss-like shuttling of protons for delocalized and rapid transport [6]. This feature leads the solvation properties of water ions to differ from that of similarly size monovalent ions [21]. The solvent's role in mediating the separation of water ions is thus different from that of simple monovalent salts. It has been shown with *ab-initio* simulation that the dissociation of bound water ions, a process known as autoionization, requires the well-timed coordination of solvent induced electric field fluctuations and the making/breaking of hydrogen bonds. These correlated fluctuations result in a near-spontaneous relocation of a proton from a neutral water molecule to a newly-formed hydronium ($H_3O^+$)[6, 12], leaving behind a negatively charged hydroxide ion ($OH^-$).

Because water's mechanistic role in charge separation differs between classical ions and water ions, changes in the aqueous environment can thus yield ion-specific effects. In this manuscript, we consider specifically the environment changes that emerge at the interface between liquid water and an extended planar platinum electrode. At such an interface, strong water-metal interactions lead to the formation of an electrode-adsorbed water monolayer [22–25]. Snapshots taken along charge separation trajectories near the electrode interface are shown in Fig. 1(A,B). This monolayer has been shown to exhibit molecular relaxation dynamics that are orders of magnitude slower than that of the bulk liquid [8]. This slowly evolving water monolayer affects the structure and dynamics of the adjacent liquid [26]. For pure water systems this effect is subtle in comparison to the dramatic slowdown of the monolayer itself, however, ions at this liquid interface can incorporate monolayer waters into their solvation shell and thus couple directly to the slow dynamics of adsorbed monolayer.

### Simulating rare events in heterogeneous environments

To study ion dynamics at the aqueous electrode interface we perform atomistic simulations of liquid water in contact with the (111) surface of an extended platinum electrode. We utilize two different model systems: The first is designed to study the dissociation of a classical ion pair, $Na^+$ and $I^-$, and the second is designed to study the recombination of water ions, $H_3O^+$ and $OH^-$. In the first model system we describe the aqueous solution using classical force fields. Specifically, we describe water using the SPC/E model [27], and we describe the ions similarly, as spherically symmetric point charge particles. This efficient combination has been demonstrated to accurately reproduce the molecular structure and dynamics of liquid water as well as experimental measures of ion hydration and mobility. Unfortunately, this classical nondissociative model is inadequate to describe the dynamics of water ions, whose transport is facilitated by the making and breaking of covalent OH bonds. Thus, in the second model system we use an *ab-initio* model of water based on density functional theory (DFT)[6].

For both model systems the electrode and its interaction with water molecules is described following the model of Siepmann and Sprik [28]. In this model each electrode atom includes a fluctuating partial charge that varies in order to maintain a constant potential condition. In this way, fluctuations in the charge distribution of the aqueous environment induce corresponding fluctuations in the electronic polarization of the electrode. These fluctuations mimic the electrostatic contributions of image charges and are consistent with a textbook description of a metal in that they obey the Johnson-Nyquest relation[29]. Other electronic degrees of freedom, such as those that determine the details of water-platinum binding, are described implicitly using empirical interaction potentials. This electrode model captures important aspects of molecular physics that are unique to electrode interfaces, such as electronic polarization and site-specific water adsorption, at a fraction of the computational cost associated with an *ab-initio* description of the metal.

We analyze charge separation in terms of both thermodynamics and kinetics. We quantify the thermodynamics by computing the free energy along the microscopic coordinate the characterizes transitions between the bound and charge separated states. For the classical ion system, we compute this free energy using standard umbrella sampling techniques[30] and we compute the dissociation rate constant from the flux-flux correlation function[31]. These methods are direct and accurate but they are too computationally demanding to be applied to the water ion system, which relies on relatively expensive *ab-initio* molecular dynamics simulations. Thus, for the system with water ions we use our limited ensemble of trajectories to construct of a Markov State Model (MSM) and we analyze this model to infer the thermodynamics of charge recombination[32]. To do this we discretize the charge separation order parameter, as defined in the SI, into a series of meta-basins and then we use these meta-basins as a basis for a MSM. The MSM is parameterized by specifying the transition rates between meta-basins, which we compute directly from our trajectory ensemble. This MSM framework allows us to infer the thermodynamic consequences associated with subtle difference between simulations carrier out in the bulk and at the electrode interface. We analyze the kinetics of water ion pairs by analyzing the statistics of recombination times as derived from simulation data and an associated model of charge recombination dynamics.

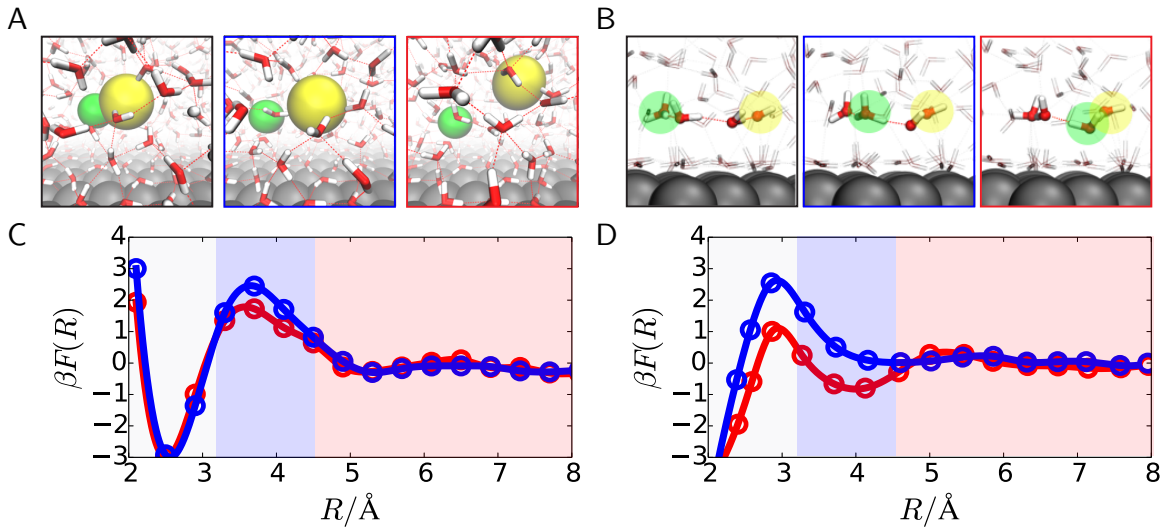

FIG. 1. Panels (A) and (B) contain typical snapshots going from a recombined state (black-bordered panel) to a dissociated state (red-bordered panel) for classical ions and water ions respectively. The positive ion is highlighted in green and the negatively charged ion is highlighted in yellow. Panels (C) and (D) contain plots of the free energy profile as a function of $R$, the inter-ionic separation for classical ions and water ions respectively. In both panels the data plotted in red corresponds to ions in the bulk and the data plotted in blue corresponds to ions at the electrode interface.

**Thermodynamics of interfacial charge separation**

The thermodynamics of aqueous charge separation can be evaluated by considering the free energy, or reversible work, to separate two oppositely charged particles in solution, such as plotted in Fig. 1(C,D). As this figure illustrates, the free energy to separate water ions and the free energy to separate classical ions bear a similar general structure. That is, both systems exhibit a stable free energy basin at small inter-ionic spacing, which we identify as the bound state, and a plateau-like region at larger inter-ionic spacing, which we identify as the charge-separated state. In both cases, these two states are separated by a transition region that includes a small ($\sim 2 k_{\mathrm{B}} T$) free energy barrier in the direction of association.

The thermodynamics of charge separation depends significantly on the solvation free energy of the ion pair. This free energy is shaped by the properties of the aqueous environment through two primary contributions. The first contribution is due to differences in the solvation free energy between the bound and charge separated states. This free energy difference controls the relative heights of the bound basin and the charge separated plateau. The second contribution is due to distortions in the ionic solvation shell that arise in response to the changes in inter-ionic distance that occur during transitions between the bound and charge separated states. These distortions determine characteristics such as the size and shape of the free energy barrier.

As indicated in Fig. 1, aqueous contributions to the free energy depend on both the identity of the charged particles and also on the details of the aqueous environment. We focus on the latter by comparing the charge separation free energy for ions at the electrode interface to that of ions in the bulk liquid. We observe that near an electrode both water ions and classical ions face an increased free energy barrier for recombination. However, this barrier increase is larger for water ions than it is for classical ions. This differing influence of the electrode on the charge separation free energy of classical ions and water ions reflects water's differing role in the charge separation mechanism for these two systems. That is, the electrode's influence on charge separation is determined by water's specific role in the microscopic process. To understand this more thoroughly we now describe the mechanistic details for each case separately, focusing specifically on the indirect role of the electrode on the water-mediated aspects of charge separation.

**The mechanism of charge separation for simple ions**

The mechanism of ionic charge separation involves an interplay between ion and water dynamics. To understand this interplay and how it is affected by electrode interface, we first identify the reaction coordinate that encodes the microscopic details that are relevant to the dynamical process. An appropriate reaction coordinate must be capable of (**1**) distinguishing between the bound and charge separated states, and (**2**) properly characterizing the transition state ensemble (TSE), which are those configurations that have equal probability of committing to either the bound or charge separated state[1]. The one-dimensional coordinate of Fig. 1 accomplishes the former but not the latter. It omits the fundamental role of the solvation environment in facilitating transitions between the bound and charge separated states.

The configuration of water molecules in the system determine the electrostatic environment of an ion pair. Collective rearrangements of water molecules control the electrostatic variations that drive the process of ionic charge separation. We can quantify these variations by computing the solvent-induced electrostatic potential, also known as the Madelung potential,

$$\psi_\pm = \frac{q_\pm}{4\pi\epsilon_0} \sum_i^{N_{\mathrm{H_2O}}} \frac{q_i}{|\mathbf{r}_i - \mathbf{r}_\pm|}, \qquad (1)$$

where the summation is taken over all of the atoms that belong to water molecules, $q_i$ and $\mathbf{r}_i$ is the charge and position of the $i^{th}$ water atom, $q_+$ and $\mathbf{r}_+$ are the charge and position of the Na$^+$ ion, and $\epsilon_0$ is the vacuum permittivity. This quantity reports directly on water's collective contribution to the electrostatic environment of the Na$^+$ ion, and similarly the analogous quantity for I$^-$, $\psi_-$, reports on collective water fluctuations around it. Their sum, $\psi = \psi_+ + \psi_-$, is thus the total electrostatic potential acting on the ionic pair, from the surrounding solvent, for convenience we report it in units of $1/\beta = k_\mathrm{B}T$.

The quantity $\psi$ exhibits different statistics when the ions are in their bound state compared to when then when they are independently solvated. This can be seen in Fig. 2, which contains a plot of the charge separation free energy as a function of $\psi$ and $R$. When resolved in these coordinates the bound and charge separated states are connected via a pathway that is mutually elongated in both $\psi$ and $R$, clearly indicating the importance of collective water fluctuations in ionic charge separation.

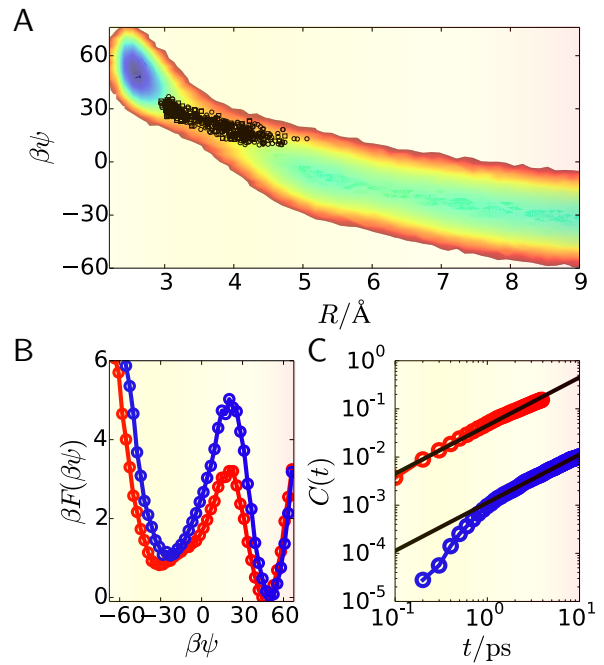

FIG. 2. Mechanism of ion pair dissociation for NaI. (A) Free energy as a function of inter-ionic separation, $R$, and the total Madelung potential, $\psi$, shown in the contour plot. Members of the TSE are shown as black points on this surface. (B) Free energy as a function of the Madelung potential for the bulk (red) and interface (blue). (C) Flux-flux correlation function for the classical ions for the bulk (red) and interface (blue), with linear fits shown in black.

Furthermore, members of the TSE harvested with transition path sampling are distributed along this free energy path, suggesting that the two-dimensional coordinate of $(\psi, R)$ satisfies the criteria to be a suitable reaction coordinate. We have confirmed this suitability by performing commitor analysis across members of the TSE, (see SI).

We isolate water's contribution to the charge separation process by projecting the free energy onto the coordinate $\psi$. The result, plotted in Fig. 2B, reveals that the free energy cost for solvent reorganization is larger at the electrode interface, by about $1.5k_\mathrm{B}T$, than it is in the bulk liquid – a consequence of constraints imposed on the liquid by the presence of the electrode. This difference in barrier height is expected to have consequences for the relative dissociation kinetics. Based on transition state theory, identical systems with a barrier height difference of $1.5k_\mathrm{B}T$ are predicted to have rates that differ by a factor of $\exp(1.5) \approx 4.5$. That is, based on barrier height alone we expect $k_\mathrm{elec} \approx 0.2k_\mathrm{bulk}$. However, when dissociation rates are computed directly, via the flux-flux correlation function shown in Fig. 2C, we find that $k_\mathrm{bulk} \approx 0.02k_\mathrm{elec}$, a full order of magnitude different. The resolution to this apparent discrepancy is that the bulk environment is not identical to that of the electrode interface, and that these differences affect the flux over the barrier in the free energy surface.

Water dynamics are more sluggish at the electrode interface than they are in the bulk liquid [8]. Strong water-electrode interactions lead to the formation of an adsorbed water monolayer and molecules within this monolayer experience slow orientational relaxation dynamics. For ions near the electrode the charge separation process couples to this slowly evolving monolayer, which significantly increases the timescale governing dynamics in $\psi$. Together, this reduced diffusion and the increase in free energy barrier height combine to yield the anomalously large difference in dissociation rate between the bulk and the electrode interface.

The influence of the electrode on the dissociation of ions is mediated by the interfacial aqueous environment. This influence is controlled by water's specific role in driving the dynamics of ion pairs and how that role is affected by the presence of the electrode. For the simple classical ions described above, electrode-induced constraints on the collective reorientations of water molecules lead to a significant slowdown in dissociation rates. The electrode's influence is fundamentally different for water ions pairs, which do not require such large collective reorientations.

**The mechanism of charge separation for water ions**

The bound state of $H_3O^+$ and $OH^-$ is simply a pair of neutral water molecules. This exceptionally stable state dissociates reluctantly and is thus difficult to simulate, even with the help of rare event sampling techniques. We therefore choose to focus on the time-reversed process of charge recombination, which is both rapid and spontaneous. The recombination process occurs through the concerted transfer of protons along a chain of hydrogen bonds that connect the $H_3O^+$ and $OH^-$ [6]. During this process a large electric field fluctuation drives the shuttling of an excess proton from the $H_3O^+$ to the $OH^-$ via the participation of two or three intermediate water molecules. An illustration of this process at the electrode interface is shown in Fig. 1B.

We begin by considering the kinetics of water ion recombination, which we quantify by computing the time, $\tau$, for initially separated water ions to reach the bound state. Since $\tau$ depends sensitively on the initial separation of the ions we only compare trajectories initialized from an ensemble of configurations that have been equilibrated with a fixed hydronium-hydroxide separation of 5Å. From the free energy function in Fig. 1D, this distance is sufficiently large that the ions are not already committed to the bound basin. Starting from larger distances affects the quantitative time to recombine, but that time is just that associated with diffusion to the barrier. Water ion pairs separated by 5Å are typically bridged by chains of three consecutive hydrogen bonds via two intermediate water molecules. These hydrogen bond chains facilitate rapid charge recombination, with average recombination times of $\tau_{\text{bulk}} = 0.35$ ps and $\tau_{\text{elec}} = 0.40$ ps for ions in the bulk liquid and at the electrode interface respectively. Notably, and unlike the case of monovalent salts, the presence of the electrode has little effect on the average recombination dynamics of water ions. Despite this, however, we observe that the electrode has a subtle effect on the statistics of $\tau$.

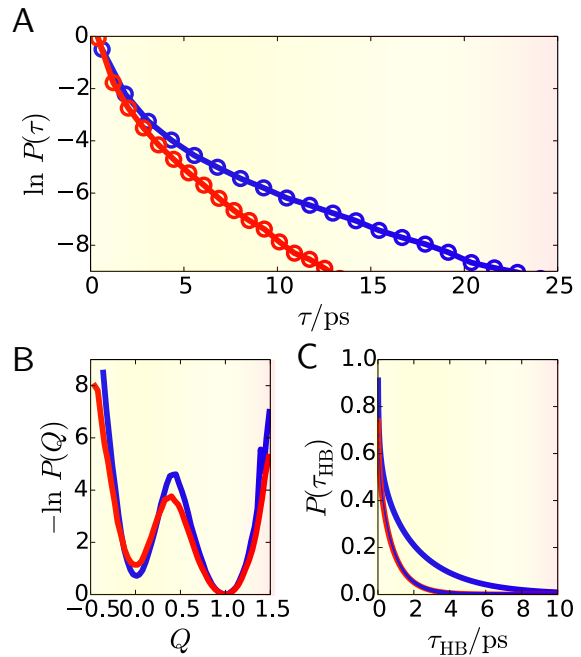

FIG. 3. Mechanism for water ion association for bulk (red) and the interface (blue). (A) The distributions of recombination times, $\tau_{KMC}$. (B) Distribution function of the net system charge. (C) Distribution of hydrogen bond lifetimes.

We quantify the electrode's effect on the recombination time statistics by computing $P(\tau)$, the probability that a given trajectory with initial ion separation of 5Å will have a recombination time of $\tau$. Comparing $P(\tau)$ for ions in the bulk and the electrode interface reveals discernible differences in the large-$\tau$ tails of the distributions. However, as shown in the SI, these differences are difficult to quantify due to poor statistics (our simulation data includes $\approx 200$ independent trajectories). We thus improve statistics with the use of a simplified model of water ion recombination derivable from this coarse data. This model described in more details within the SI is designed to efficiently predict the value of $\tau$ given the positions of $H_3O^+$, $OH^-$, their surrounding water molecules, and the connectivity of the hydrogen bond network. The model describes charge recombination in terms of two types of dynamic events. First, a single proton can hop along a hydrogen bond to the $OH^-$ or from the $H_3O^+$. This results in an exchange of positions of a water ion and neighboring water molecule. Second, the charges can recombine, via a concerted mechanism, through short chains of hydrogen bonds. The rate for this process is a function of the number of hydrogen bonds in the chain. We simulate charge recombination with a kinetic Monte Carlo (KMC) algorithm for fixed configurations of the hydrogen bond network. We compute the rates of the processes in the KMC model directly from our simulation data.Figure 3A contains a plot of $P(\tau)$ computed by applying our KMC model across the ensemble of initial conditions used in our *ab-initio* simulations. We observe that recombination in the bulk and at the electrode interface have similar mean behavior but differ in the tails of their recombination time distributions. Specifically, it is more likely that an ion pair at the electrode interface has an abnormally long recombination time. This large difference involves very low probability events and thus does not manifest itself in the mean behavior. We attribute the difference in these curves to subtle differences in the topology of the hydrogen bond network between the bulk and the interface. Proton dynamics in our model are constrained by the hydrogen bond network and thus especially sensitive to the details of network structure. Specifically, the interfacial hydrogen bond network is distorted along the plane of the interface. This distortion slightly reduces the relative number of hydrogen bond chains along which recombination can occur. This reduction is balanced by an increase in the probability that a proton transfer will increase the separation between the ions, thus leading to an increase in recombination time.

The electronic and nuclear rearrangements that accompany water ion recombination are driven by electrostatic fluctuations of the aqueous environment. These fluctuations are reflected in the distribution of electronic charge in the system, which is different for the bound and charge separated states. One way to quantify this difference is to compute the imbalance of electronic charge between the hydroxide and the hydronium species, as given by,

$$Q = \frac{1}{2}\left[\sum_{i \in \mathrm{H_3O+}} q_i - \sum_{j \in \mathrm{OH^-}} q_j\right], \qquad (2)$$

where the two summations are taken over all the atoms belonging to the hydronium or hydroxide ion respectively, and $q_i$ is the Mulliken charge on the $i$th atom. We define the hydroxide and hydronium based on nuclear coordinates by first associating each hydrogen with its nearest oxygen, as described in the SI, and then identifying an oxygen with only one associated hydrogen as a hydroxide center and an oxygen with three associated hydrogens as a hydronium center. These water ion centers and their associated hydrogens are thus included in the summations in Eq. 2 [33]. This continuously varying quantity is formulated so that $Q = 1$ when the hydronium and hydroxide are fully charged and $Q = 0$ when the two species are charge neutral. By analyzing the statistics of $Q$, sampled over an ensemble of recombining trajectories, we can compute the free energy for charge recombination.

Fig. 3B shows the probability distributions for $Q$ for water ion recombination in the bulk and at the electrode interface, computed directly from the ensemble of recombination trajectories [34]. These profiles reveal that the barrier height to recombination is slightly larger at the electrode interface than it is in the bulk liquid. This barrier height difference is approximately $1.5 k_\mathrm{B} T$, similar to that found in the case of the classical ions described above. However, the kinetic consequences of this observation are not apparent when comparing the recombination times, which are nearly identical near and away from the electrode interface. We can thus conclude that the kinetic effect of an increased barrier height is compensated by an increase in flux along the coordinate $Q$.

We explain the molecular origins of the enhanced flux in $Q$ in terms of the equilibrium dynamics of hydrogen bonding. The concerted proton hops that drive recombination are directed along chains of multiple hydrogen bonds. This process is therefore contingent upon the stability of these chains. Fluctuations that destabilize or sever hydrogen bond chains will have a negative effect on recombination. Along a liquid water interface such fluctuations are suppressed due to geometric constraints that lead to more stable and longer lived hydrogen bonds [35, 36]. The consequences of these constraints can be quantified by computing the distribution of hydrogen bond lifetimes, $P(\tau_{\mathrm{HB}})$. In Fig. 3C we see that the hydrogen bond lifetimes are larger at the electrode interface than they are in the bulk. Hydrogen bond chains at the interface are thus longer lived which allows more time for the environmental fluctuations to induce the electronic reorganization associated with charge recombination.

## IMPLICATIONS FOR ELECTROCHEMISTRY

The aqueous environment of an electrode interface differs from that of the bulk liquid. These differences affect the microscopic processes that underlie many electrochemical applications. In this manuscript we have highlighted specifically the microscopic process of ionic charge separation and shown that the electrode's influence can depend significantly on the identity of the ionic species. This dependence is controlled by water's specific role in mediating ion pair dynamics. We find that for ions comprising monovalent salts, charge separation is slowed near an electrode interface by over an order of magnitude, however, the analogous process for water ions is barely affected. Understanding and accounting for these effects, and their dependence on ion type, is important for the design of electrochemical systems.

Importantly, the dominant influence of the electrode on the separation of ion pairs is dynamical. Electrode-water interactions alter the equilibrium properties of the aqueous environment, such as the timescales that govern molecular fluctuations, which couple to the microscopic dynamics of ionic charge separation. The details of the electrode-water interactions thus control these dynamical effects. In these results we utilize a platinum (111) electrode, however, an electrode made of a different metal, such as gold or copper, or with a different surface geometry and water binding energy may have a different influence on the dynamics of charge separation[22]. Notably, these dynamical effects are not apparent in the thermodynamics and are thus difficult to predict without a model that explicitly describes the interplay between ions, their solvation shell, and the aqueous dynamics at the electrode interface. Traditional continuum models are thus insufficient to predict these significant interfacial effects.

### Methods

All simulations were performed in an ensemble with constant number of particles, volume, temperature, and applied electrode potential. The classical simulations were set up with simulation cell sizes, $3 \times 3 \times 6$ nm, and the *ab-initio* simulations with cell sizes, $1 \times 1 \times 2$ nm. The later used the PBE functional with D3 dispersion correction and GTH pseudopotentials. More details can be found in the SI.

### ACKNOWLEDGMENTS

This work was supported by the National Science Foundation under CHE-1654415 (A.P.W.) and the MIT Department of Chemistry through junior faculty funds (J.A.K. and A.P.W.). D.T.L. was initially supported by the Princeton Center for Theoretical Science and later by University of California, Berkeley, College of Chemistry. We thank Dorothea Golze for assistance with setting up the *ab-initio* simulations.

SUPPORTING INFORMATION

I. CLASSICAL IONS

A. Classical Simulations

The system simulated consists of a slab of water in contact with a metal surface on each side. The metal surface consists of three layers of atoms, totaling nearly 500 particles, held fixed in a face centered cubic lattice with a Pt spacing of 3.92 Å and the 111 facet exposed to the solution. A slab of water nearly 40 Å thick was placed in contact with the metal, and the dynamics of the nearly 1800 molecules were propagated using a langevin thermostat, with SETTLE[1] imposing bond and angle constraints for the water as implemented in LAMMPS [2]. All simulations were run at 298 K. Interactions between the water molecules are computed from the SPC/E potential [3]. The water-metal potential is modeled following Siepmann and Sprik [4] where the platinum water interaction is a sum of two and three body terms, parameterized to get the correct value of the adsorption energy and ground state geometry as determined by quantum chemical calculations. Additionally, to model the polarizable metal surface each atom carries a Gaussian charge of fixed width but variable amplitude, which is updated at each timestep by minimizing the energy of the slab subject to a constraint of equal potential across the conductor. Periodic boundary conditions are employed in the plane parallel to the surface. Ewald summations appropriate for mixed point and Gaussian charge densities were employed[5]. A more thorough description of this model can be found elsewhere [4, 6]. The Na$^+$ and I$^-$ ions were modeled with point charges and Lennard Jones potentials[7]. A reduced charge was used on the I$^-$ of $-0.75$ so that the ion pair has a favorable free energy of adsorption to the interface[8].

B. Free energy calculations

In order to compute the free energy as a function of interionic separation, $R$, we have preformed umbrella sampling in $R$ and $z$, the mass weighted distance away from the electrode of the ionic pair. Specifically, we apply a bias potential of the form

$$\Delta U(R, z) = k((z(t) - z^*)^2 + (R(t) - R^*)^2)/2$$

where $k$ is a spring constant equal to 13.5 $k_\mathrm{B}T/\text{Å}^2$ and $z^*$ and $R^*$ are umbrella[9] minimum, chosen on a uniform grid of 100 point between $z = \{8:15\}\text{Å}$, and $R = \{2,10\}\text{Å}$. Each trajectory was integrated for 2 ns following 50 ps of equilibration. Unbiasing and reweighting were accomplished with the multistate Bennet acceptance ratio[10]. In this way the joint free energy, $F(R,z)$, was computed from

$$\beta \Delta F(R,z) = -\ln P(R,z)$$

which is shown in SI 1. Shown in Fig. 1C of the main text are conditional distributions, or free energies, at select values of $z = 9$, and 15, or $F(R|z)$ with $z$ fixed.

Because the madelung potential, $\psi$ and the interionic distance, $R$, are highly correlated thermodynamically, we have used the indirect umbrella sampling method[11] to compute the joint free energy $F(R,z,\psi)$ from umbrella sampling simulations used to construct $F(R,z)$. This is constructed by

$$\beta \Delta F(R,z,\psi) = -\ln P(R,z,\psi)$$

without additional sampling. To compute $F(\psi,R)$ we then compute the marginal distribution as

$$F(\psi,R) = -kT \ln \int dz e^{-\beta F(R,z,\psi)}$$

which is shown in Fig. 2 of the main text. Alternatively we can construct the free energy for changing $\psi$ at fixed $z$, by marginalizing over $R$ as,

$$F(\psi|z') = -kT \ln \int dR e^{-\beta F(R,z=z',\psi)}$$

which is plotted in Fig. 2B of the main text.

### C. Transition path sampling

To analyze the kinetics of ion pair dissociation, we have used transition path sampling[12]. Specifically, we have studied ensembles of trajectories conditioned on starting with inter-ionic distances, $R < 3.5\,\text{Å}$, and ending with $R > 4.5\,\text{Å}$ over a trajectory 10 ps in length. The starting and ending conditions were chosen to adequately delineate configurations committed to the bound and free states, and the trajectory length was chosen in order to ensure that ions had sufficient time to cross the barrier. Standard shooting and shifting moves were used with a ratio of 1:3. Shooting moves were attempted uniformly along the trajectory and shifting moves were attempted uniformly along the first and last 1/3. Because the trajectories were generated from stochastic dynamics, shooting moves were accomplished by simply reintegrating the trajectories with different randomly drawn noises[13]. This procedure resulted in trajectories decorrelating within around 3 moves, as measured by trajectory autocorrelation functions of the average $R$ and $z$. In total, 1000 independent trajectories were generated.

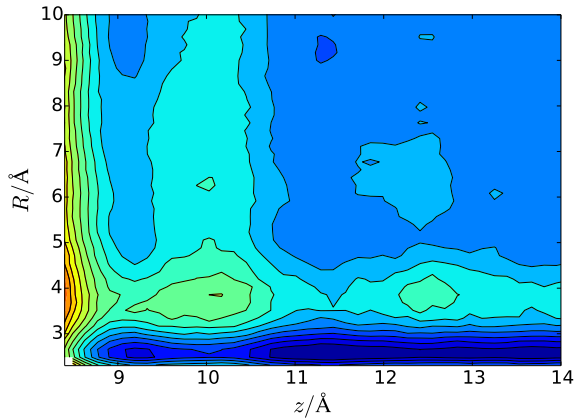

FIG. 1. Joint free energy for moving the ion center of mass towards or away from the electrode, $z$, and separating the ion pair, $R$.

To generate members of the transition state ensemble, TSE, we used trajectories of length 10 ps and average over 10 separate realizations of velocities taken from the Maxwell-Boltzmann distribution so as to compute a commitment probability, $p_B$ for a configuration dissociating. If this estimate of $p_B$ falls within a 90% confidence interval of 0.5, we say it is is a member of the TSE. Members of the TSE are projected into the $R, \psi$ plane in Fig. 2A. In order to evaluate $\psi$ as a relevant dynamical coordinate, we have computed the distribution of commitment probabilities constrained to the saddle point in $\psi$, which is at $\beta \psi = 15$, and compared that distribution to that constrained to $R = 4$. Each estimate of the commitment probability was computed from 10 independent trajectories, and around 500 configurations were generated for each coordinate from constrained equilibrium sampling. Figure 2 shows the results of these calculations, with the distribution constrained to $\beta \psi = 15$ yielding a peaked distribution around $p_B = 0.5$ and the distribution constrained to $R = 4$ yielding a bimodal distribution around $p_B = 0.0$ and $p_B = 1$. This is strong evidence for $\psi$ being the relevant dynamical variable for charge separation[14].

Rate constants were evaluated by computing the side-side correlation function[15]. Specifically the correlation function,

$$C(t) = \frac{\langle h_a(0) h_b(t) \rangle}{\langle h_a(0) \rangle}$$

where,

$$h_a(t) = \begin{cases} 1 \text{ if } R(t) < 3.5\,\text{Å} \\ 0 \text{ else} \end{cases}$$

and

$$h_b(t) = \begin{cases} 1 \text{ if } R(t) > 4.5\,\text{Å} \\ 0 \text{ else} \end{cases}$$

are the basin definitions used in the transition path sampling calculations. The long time slope of this function is the rate of dissociation, which was computed for 1000 independent trajectories initiated from constrained equilibrium sampling both near or away from the interface. These functions were reported in figure Fig. 2C of the main text.

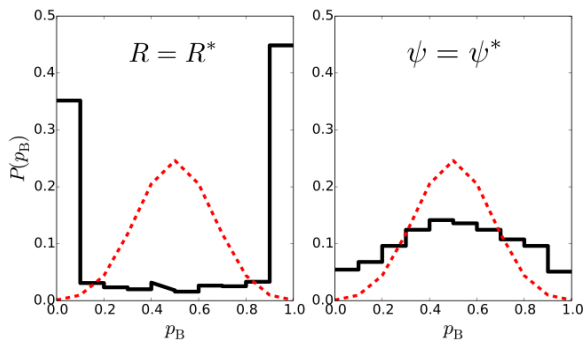

FIG. 2. Conditional commitment probability $p_B$, evaluated with TPS (black), for $R = R^* = 4\text{Å}$ (left) and $\beta\psi = \beta\psi^* = 15$ (right) compared to a Bernoulli distribution with 10 trials (red).

## II. WATER IONS

**DFTMD simulations**

To simulate water ion recombination Density Functional Theory based Molecular Dynamics (DFTMD) simulations were performed. The Gaussian Plane Wave (GPW) imple-mentation [16] in CP2K [17] was used with a DZVP Gaussian basis set [18] and a plane wave cut off at 280 Ry for the bulk simulations and 300 Ry for the interface simulations. The core electrons are described with GTH pseudopotentials [19, 20]. The calculations were done at the PBE level [21]. Grimme's D3 dispersion corrections [22] were used for the interface calculations. The fixed metal surface was modeled with the same lattice constant of 3.92 Å, with the 111 facet exposed to water as described above. The IC-QM/MM method [23] was used to model Siepman-Sprik metal and the same empirical potential was used for the adsorption of water molecules as for the monovalent ions [4]. The ensemble was simulated using a time step of 0.5 fs with a target temperature of 330 K, which was controlled using the canonical sampling velocity rescaling thermostat [24].

The bulk systems consisted of 32 and 64 water molecules inside cubes of length 9.85 Å and 12.41 Å respectively. An orthorhombic supercell was constructed for the interface system by taking a subset of atoms from a larger equilibrated classical constant potential simulation by only including atoms in a box 13.90 × 14.44 × 22.0 Å and then adding a 20 Å vacuum buffer. The number of water molecules in the system were: 106, 98 and 96. Periodic boundary conditions were applied in all three dimensions.

Initial conditions were taken from equilibrated classical molecular dynamics simulations of the form described above, followed by 10 ps of DFTMD equilibration. A proton was then removed from a water molecule and attached to another water molecule a distance $R$ away in order to create a pair of charged water ions. Constraints to impose a fixed proton coordination number were then applied and the system was equilibrated for 10 ps. Initial configurations for recombination trajectories were taken every 100 fs, after the initial 6 ps. The recombination trajectories were run with no constraints and were terminated upon recombination to neutral water molecules. This initialization procedure was redone independently three additional times in the bulk and twice at the interface.

**Identifying water ions**

To identify the $H_3O^+$ and $OH^-$ species in the DFTMD simulation the coordination of the hydrogen atom with an oxygen atom is calculated as:

$$n_{\text{OH}}(r) = (1 - (r/r_0)^{16})/(1 - (r/r_0)^{56})$$

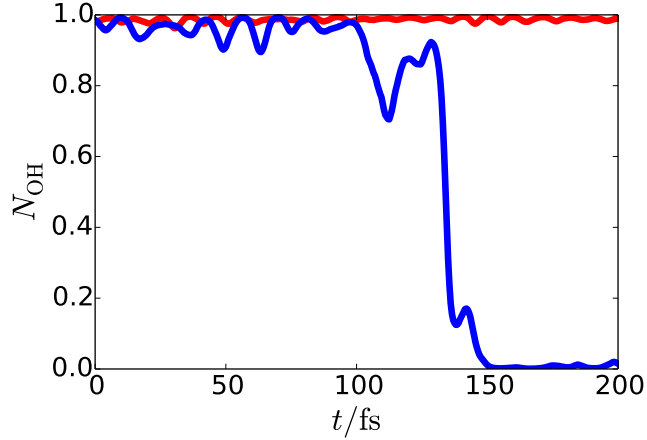

FIG. 3. The quantity $n_{OH}$ decays smoothly from 1 to 0 as an O-H bond breaks (blue) whilst remaining approximately 1 for an intact water molecule (red).

from reference [25], where $r$ is the distance between the H atom and the O atom and $r_0 = 1.32 \text{Å}$. Calculating $n_{\text{OH}}$ for every oxygen atom with each hydrogen atom and taking the maximum value allows for the assignment of each hydrogen atom with one oxygen atom. The oxygen atom that coordinates with three hydrogen atoms is the $H_3O^+$ center and the oxygen atom that only coordinates with one hydrogen atom is the $OH^-$. In some instances the distances can be very delocalised and more than one $H_3O^+$ and $OH^-$ centers are identified. In these cases a cluster of ions can be identified *i.e.*, a hydronium with the structure $[H_3O - OH - H_3O]^+$ or a hydroxide with the structure $[OH - H_3O - OH]^-$. In these cases we define the water ion as the central oxygen and the three or one nearest hydrogens for hydronium or hydroxide respectively.

An example of how the quantity $n_{\text{OH}}$ decays from 1 to 0 as the O-H bond breaks is shown in figure 3. This is compared with a an O-H bond that vibrates whilst remaining in tact.

**Recombination time from DFTMD simulations**

The number of ions in the simulation is obtained using the same order parameter as in reference [25] where the number of ions:

$$N_{\text{ions}} = \sum_{i=1}^{N_O} (n_i - 2)^2,$$

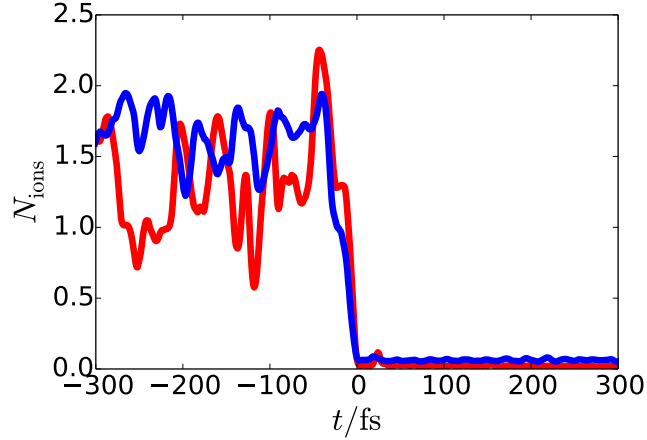

FIG. 4. The recombination of water ions is followed by analyzing the number of ions in the simulation as seen for a typical trajectory. The recombination time is shifted to $t = 0$.

where $N_O$ is the total number of oxygen atoms in the system and $n_i$ indicates the number of hydrogens coordinated to the $i$th oxygen,

$$n_i(r) = \sum_{j=1}^{N_\text{H}} (1 - (r_{ij}/r_0)^{16})/(1 - (r_{ij}/r_0)^{56}),$$

where the summation is taken over all $N_\text{H}$ hydrogens in the system, $r_{ij}$ is the distance between oxygen $i$ and hydrogen $j$. The recombination time for a given simulation is obtained as the first instance in time where $N_\text{ions}$ approximately equals 0. An example of this shown in figure 4.

Figure 5 shows the DFTMD data for recombination times of 200 simulations where the initial ion-ion distances were fixed at 5 Å. The noisiness of this data motivates the use of the KMC model.

**Hydrogen bond lifetime**

The hydrogen bond network is analyzed using a standard definition where the O-O length must be less than $3.5 Å$ and the O-H-O angle must be less than $30°$. The hydrogen bond lifetime $\tau_\text{HB}$ is obtained by identifying the hydrogen bonds at $t = 0$, using this hydrogen bond definitions, and finding the time that the hydrogen bond breaks.

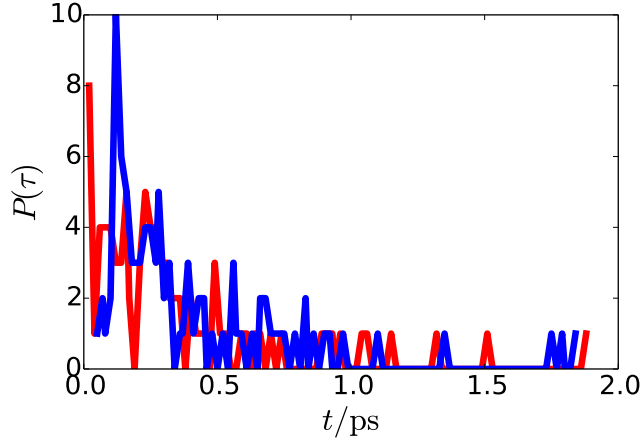

FIG. 5. The probability of recombination time $\tau$ for the bulk (red) and the electrode (blue) for 200 simulations with ion-ion initial distances at 5 Å.

**Free energy calculations**

To compute the free energy profile from a Markov state model the water ions are identified, as outlined above, and the inter-ionic distance $R$ is calculated based on the oxygen centers. Markov states are defined by discretizing $R$ in bins 0.2 Å width and their transition probabilities are constructed from:

$$M_{R,R'} = \frac{\sum_k N_{R,R'}^k(\Delta t)}{\sum_k n_R^k}$$

where $N_{R,R'}^k$ is the number of transitions between states $R$ and $R'$ over a lag time $\Delta t$ from the $k$th trajectory, and $n_R^k$ is the total amount of time spent in state $R$ in trajectory $k$. For a fixed discretization of $R$, the lag time was chosen to be 50fs which ensures that the eigenspectrum of transition matrix shows significant gap, ensuring the transitions are Markovian.

Because the our trajectories are generated without a bias, the transition probabilities obey local detailed balance, $M_{R,R'}/M_{R',R} = p_{R'}/p_R = \exp[-\beta(E(R') - E(R))]$. The transition matrix satisfies an eigenvalue equation of the form,

$$M(R, R')p(R) = \lambda p(R)$$

the $\lambda$ are the eigenvalues. By conservation of norm, the largest eigenvalue, $\max(\lambda) = 0$, and its eigenvector is the stationary distribution. The free energy profile is given by,

$$F(R) = -k_B T \ln \sum_{R'} M(R, R') p(R').$$

which is plotted in the main text. The accuracy of this estimate of the free energy relies on locally sampling the transition probabilities in proportion to their Boltzmann weights, which requires both forward and backward transitions are well sampled. While the recombination trajectories are able to provide reasonable estimates of $F(R)$ along a large range of $R$, we are unable to report on the stability of the recombined state as the backwards transition out of the bound state is undersampled.

**Kinetic monte carlo**

A Kinetic Monte Carlo model was implemented to obtain the distribution of recombination rates. The model takes static configurations from the DFT trajectories. The identification of the hydronium and hydroxide ions and analysis of the hydrogen bond network were done using the definitions outlined above. A proton can hop to the $OH^-$ or from the $H_3O^+$ with the structural diffusion rate parameter $k$. Spontaneous charge recombination can occur through any hydrogen bond chain that connects the $H_3O^+$ and $OH^-$. The recombination rate depends on the number of hydrogen bonds in the chain and is given by $k'_n$ for a chain with $n$ hydrogen bonds. Species are also considered recombined if they reside on the same oxygen atom. Values of $k$ and $k'$ were optimized using the maximum likelihood estimation with the DFTMD simulations. The optimal values for the bulk were $k = 1/1150$ fs, $k'_{n \leq 3} = 1/325$ fs, and $k'_{n>3} = 0$. The optimal values for the interface were $k = 1/1075$ fs, $k'_{n<=3} = 1/500$ fs, and $k'_{n>3} = 0$. A screened Coulombic attraction has been included in the structural diffusion parameter by using the experimental value of the dielectric constant of water. We generate statistics by performing KMC simulations across an ensemble of starting configuration from our *ab-initio* study.

---